\documentstyle[aps,psfig]{revtex}
\begin{document}
\def\be{\begin{equation}}
\def\ee{\end{equation}}
\def\bea{\begin{eqnarray}}
\def\eea{\end{eqnarray}}
\renewcommand{\thefootnote}{\fnsymbol{footnote}}

\title{Wegner-Houghton equation and derivative expansion}

\author 
{Alfio Bonanno \footnote{bonanno@axpfct.ct.infn.it}$^{(1,2)}$,
Vincenzo Branchina \footnote{branchina@sbgvax.in2p3.fr}$^{(3)}$,
Herve Mohrbach \footnote{mohrbach@sbgvax.in2p3.fr}$^{(4)}$ 
and Dario Zappal\`a \footnote{dario.zappala@ct.infn.it} $^{(2,5)}$}

\address{$^{(1)}$Istituto di Astronomia, Universit\`a di Catania,
viale Andrea Doria 6, 95125, Catania, Italy\\
$^{(2)}$ INFN,Sezione di Catania, corso Italia 57, 95129, Catania, Italy\\
$^{(3)}$Laboratory of Theoretical Physics , University of Strasbourg,
3,rue de l'Universite, 67000, Strasbourg France\\
$^{(4)}$LPLI-Institut de Physique, F-57070 Metz, France\\
$^{(5)}$Dipartimento di Fisica, Universit\`a di Catania, corso Italia 57, 
95129, Catania, Italy}

\date{\today}
\maketitle
\draft
\begin{abstract}

We study the derivative expansion for the effective action in the
framework of the Exact Renormalization Group  for a single
component scalar theory. By truncating the expansion to the first two terms,
the potential $U_k$ and the kinetic coefficient $Z_k$, our analysis
suggests that a set of coupled differential equations
for these two functions can be established under certain smoothness
conditions for the background field and that sharp and smooth
cut-off give the same result. 
In addition we find that, differently from
the case of the potential, a further expansion is needed to obtain the 
differential equation  for $Z_k$, according to the relative weight between
the kinetic and the potential terms. As a result, two different approximations
to the $Z_k$ equation are obtained. Finally a numerical analysis of the 
coupled equations for $U_k$ and $Z_k$ is performed at the non-gaussian
fixed point in $D<4$ dimensions to determine the anomalous dimension of 
the field. 
\end{abstract} 
\pacs{pacs 11.10.Hi , 11.10.Gh \hfill PREPRINT INFNCT/1-99}

\section{Introduction}
The dynamics of a quantum system with {\it very many} (an infinite 
number of) degrees of freedom, the subject of Quantum Field Theory, 
involves {\it very many} scales at once. As it is by now well 
understood this is the source of the main difficulties encountered 
in the study of such a system. In perturbation 
theory one of the most striking manifestation of these difficulties
is the appearance of divergent quantities. The origin of this 
nonsensical result should  probably be addressed to a wrong choice of the 
representation. In fact, unlike any problem with a finite number of 
degrees of freedom,  in the present case the choice of the 
representation is part of the solution of the dynamical problem 
due to the existence of inequivalent representations,
but up to now not very much progress has been made in this 
direction.
Moreover there are phenomena, such as the chiral symmetry breaking,
where it is necessary to resort to non perturbative methods,
since no small parameter is available for a perturbative expansion. 
Typical tools for this kind of investigations are the 
Schwinger-Dyson equations as well as the variational approximations.
Unfortunately even within these approaches divergent quantities 
are still present and, unlike the perturbative case, a 
systematic way of handling these divergences is still lacking. 
This has cast serious doubts on the validity of the results 
obtained with these methods.

From a different stand point the Renormalization Group Method, 
more precisely the Wilson formulation of the Renormalization 
Group (RG)\cite{wilson}, has been invented to handle those dynamical 
problems involving very many scales. The main idea of the method 
is to break the original problem into a series of 
subproblems each involving a more restricted scale range and to
seek for its solution by iteratively solving these subproblems.
The advantage of this method is that it is non perturbative by 
construction and that no problem of divergences is ever encountered.

There are different (more or less equivalent) ways to implement 
this program. The realization in terms of differential
equations in the momentum space\cite{wegner,polch,hasen,wett,wett1,morr},
generally named `exact Renormalization
Group equations', has been widely studied in the last years
and even employed as a tool for investigating many problems
related to phase transitions where the flow of the physical parameters  
cannot be controlled by simple perturbative calculations.

Among the various existing formulations of the exact RG equations, the 
Wegner-Houghton (WH) equation \cite{wegner} is particularly 
appealing for the transparency of its physical interpretation,
and the  most straightforward approximation to this equation,
obtained by restricting the problem to analyzing the RG evolution of the 
costant ($x$-independent) part of the action, namely the Local Potential 
Approximation (LPA), has been shown to contain many interesting features.
For instance the exactness of the LPA for $N\to \infty$ has been found in 
\cite{wegner} ($N$ is the number of scalar fields $\phi_i$), 
as well as the fixed point structure of the scalar theory
has been recovered in the LPA \cite{hasen,morris2}; furthermore, 
the typically non-perturbative
feature of convexity of the effective potential is already contained in the
LPA\cite{wett,wett2,bran1}.

In order to extract informations on the dynamics of the field, however,
it is necessary to improve the LPA by including the effects of the field 
fluctuations. The natural extension of the LPA is the obtained 
by considering a {\it Derivative Expansion} of the action, 
that is by adding in the 
general form of the action terms with higher and higher number of derivatives 
of the fields. This expansion is motivated by the expectation that,
for a sufficiently smooth background configuration,
the action should be quasi-local. A set of two coupled equations for the 
local potential $U_k(\phi)$ ($k$ is the momentum scale at which the action 
$S_k$ is defined) and the coefficient $Z_k(\phi)$ 
of the lowest order derivative term, $\partial_\mu \phi \partial ^\mu \phi$
has been deduced and studied by resorting to a smooth 
cut-off\cite{morris2,boh} that allows a weighted integration of the various 
modes appearing in the original
action. An equation for $Z_k(\phi)$ which allows to recover the lowest order 
perturbative anomalous dimension of the field in four dimensions has also 
been obtained\cite{alfdar} from the WH equation by making use of a sharp 
cut-off to integrate out the ultraviolet modes.

In this paper we critically reconsider the set of coupled equations for 
$U_k(\phi)$ and $Z_k(\phi)$ and more generally discuss the derivative 
expansion in the framework of the WH equation.
In particular, after a brief review in Sect.II of the WH method and the 
derivation of the LPA, in Sect. III motivated by the result 
of \cite{alfdar} we check the reliability of the equations derived in that 
paper and evaluate the anomalous dimension of the $g_3\phi^3$ theory in 
$D=6$ dimensions  to the order $O(g_3^2)$ and, afterwards, the anomalous 
dimension of the scalar theory at the Wilson-Fisher (WF) fixed point
in $D<4$ dimensions. 

In Sect. IV we move to a critical revision of the derivation 
of this system of coupled differential equations for $U_k$ and $Z_k$.
This gives us the possibility of reconsidering the longstanding problem
which affects the implementation of the derivative expansion of the
WH equation. As soon as a non-constant field is considered,
the differential equation for $Z_k(\phi)$ (and, more generally, the equations 
for the higher order coefficients in the derivative expansion), is affected 
by the presence of some non-analytical terms that apparently spoil 
the very differential nature of the equation. The origin of these
terms is analysed in detail and the derivation of the 
equation for $Z_k$ in the momentum space is presented. 
We shall see that under a specific assumption of smoothness of 
the background field,
it is possible to derive the differential equation and that the dangerous
terms are actually negligible. The fate of these non-analytical terms when 
the sharp cut-off is replaced with a smooth one is also discussed 
and we shall show that for a sufficiently slow fluctuating background field, 
both sharp and smooth cut-off produce the same equations. 
In addition we show that 
the assuption about the relative weight between kinetic and potential
terms in the action is crucial in the derivation itself of the $Z_k$ equation.
Our conclusions are summarized in Sect.V.

\section {LPA and LOOP EXPANSION}
Let us call $S_\Lambda [\Phi]$ the action of a 
scalar field $\Phi$ which contains all the Fourier components 
$\phi_q$ such that $0<|q|<\Lambda$. For the sake of simplicity in this paper 
we shall only consider the single component scalar theory.Usually we want to 
describe a physical process as the scattering of two bosons with momenta 
$p_\mu$ such that $p << \Lambda$. As already mentioned the 
origin of the difficulties in perturbation theory is due the fact 
that we have to take into account the contributions to this 
process  coming from all the Fourier modes between $p$ and 
$\Lambda$. Actually the action $S_\Lambda [\Phi]$, often called 
the  wilsonian effective action at the scale $\Lambda$, 
is more appropriate to describe the physics of the system at 
scales  $p\sim \Lambda$. 
At a scale $p<<\Lambda$ it would seem more convenient to directly deal 
with an effective action $S_p[\phi]$ that describes the 
physical phenomena taking into account only those 
scales around $p$. The Renormalization Group Method provides this
effective action. Let us define $S_p [\phi]$, in the Euclidean
version of the theory, through the equation

\be\label{effact}
e^{-\frac{1}{\hbar}S_p[\phi]}=\int[D\zeta]
e^{-\frac{1}{\hbar}S_\Lambda[\phi+\zeta]}
\ee

\noindent 
where we have split  the original field $\Phi(x)$ into a background 
field $\phi(x)$  containing only the 
modes between zero and $p$ and a fluctuation field $\zeta(x)$ 
containing those between $p$ and $\Lambda$,

\be\label{split}
\Phi(x)=\phi(x)+\zeta(x).
\ee

$S_p[\phi]$, called the 
 wilsonian effective action at the scale $p$, contains,
within its parameters,
the effect of the interactions among the modes in the range 
$[p,\Lambda]$ and  the modes below $p$ and therefore 
it is the effective action we were looking for. 
The difficult task here is to perform the 
integration 
over the high frequency modes $\zeta$. The above integral can 
be approximated by means of the loop expansion which amounts to a series 
expansion in powers of $\hbar$. The one loop approximation 
gives the $O(\hbar)$ correction to the tree level result 
$S_p[\phi]=S_\Lambda[\phi]$ and corresponds to a gaussian 
integral over the fluctuation $\zeta$ obtained expanding 
$S_\Lambda[\phi+\zeta]$ up to $\zeta^2$\cite{jean}.  
 
We may now pose the question in a different way 
and ask how does the wilsonian effective action $S_k$ evolve 
once we integrate the modes in the shell $[k-\delta k, k ]$ 
with an infinitesimal $\delta k$ to get $S_{k-\delta k}$. 
In this case the gaussian approximation, i.e. the one loop 
result, becomes exact\cite{wegner}. The problem of the evaluation of the 
effective action  at the scale $k$ in Eq. (\ref{effact}) can then
be turned into the problem of solving a first order differential 
equation for $S_k[\phi]$ with respect to the variable $k$.
The action $S_\Lambda [\Phi]$ is the the boundary condition
for this equation.   
This is the Wegner-Houghton method. Actually the problem of 
solving this differential equation for $S_k$ is as difficult 
as to perform the integral in Eq. (\ref{effact}) and we have to 
resort sooner or later to some approximation if we want to make 
any progress. As a first step let us go back to 
Eq. (1) which, replacing $\Lambda$ with $k$ and $p$ with $k-\delta k$,
is the defining equation for $S_{k-\delta k}$ in terms of $S_k$

\be\label{sk}
e^{-\frac{1}{\hbar}S_{k-\delta k}[\phi]}=\int[D\zeta]
e^{-\frac{1}{\hbar}S_k[\phi+\zeta]}
\ee

\noindent

Our second step is to insert in the above equation an ansatz
for $S_k[\Phi]$. At this point we make the assumption that a good 
approximation to 
$S_k[\Phi]$ is provided by the derivative expansion ($D$ is the number
of dimensions)

\be\label{ansatz}
{S}_k[\Phi]=
\int d^Dx \Big [ U_k(\Phi) +  \frac{1}{2}
Z_k(\Phi)\partial_\mu\Phi~\partial_\mu\Phi 
+Y_k(\Phi)(\partial_\mu\Phi~\partial_\mu\Phi)^2+\cdots\Big ]
\ee

\noindent
By inserting Eq. (\ref{ansatz}) into (\ref{sk}) we should be able to 
derive an infinite system of coupled differential equations for 
the coefficients functions $U_k, Z_k, Y_k, \cdots$.

The lowest order approximation in the derivative expansion is the LPA. 
Replacing in Eq. (\ref{ansatz}) $Z_k=1, Y_k=0,
\cdots$, and then considering a constant background field $\phi(x)=\phi_0$ 
we get from   Eq. (\ref{sk}) an evolution equation for $U_k(\phi_0)$ only

\be\label{poteq}
k\frac{\partial}{\partial k}U_k(\phi_0)=-
\frac{\hbar k^D N_D}{ 2}ln(k^2+U_k^{''}(\phi_0)).
\ee
\noindent
Here the $'$ means derivative with respect to $\phi_0$ and the result of the
angular integration is $N_D=2/((4\pi)^{D/2} \Gamma(D/2))$.

This equation has been found again and again 
(see for instance \cite{nicol,hasen})
and the consensus on this equation is unanimous.

The equation above is a non perturbative evolution equation 
for $U_k$. Suppose that  the potential $U_k(\phi_0)$
has a polynomial expansion and for the sake of simplicity let us require the 
$Z(2)$ symmetry $\phi_0 \to -\phi_0$

\be\label{ukappa}
U_k(\phi_0)= g_0(k) +\frac{1}{2} g_2(k)\phi_0^2
+\frac{1}{4!} g_4(k)\phi_0^4 +\frac{1}{6!} g_6(k)\phi_0^6
+\frac{1}{8!} g_8(k)\phi_0^8 +\cdots
\ee

\noindent
By taking $2n$ times  ($n=0,1,2,\cdots$ ) the derivatives of
Eq. (\ref {poteq}) with respect to $\phi_0$ at $\phi_0=0$ we get an infinite 
system of coupled equations for the coupling constants 
$g_{2n}(k)$:

\bea\label{set}
&&k\frac{\partial}{\partial k}g_2(k)=
-\frac{\hbar k^D N_D}{2}
\frac{g_4(k)}{k^2+g_2(k)}\nonumber\\
&&k\frac{\partial}{\partial k}g_4(k)=
-\frac{\hbar k^D N_D}{2}\Biggl[\frac{g_6(k)}{k^2+g_2(k)}
-3\frac{g_4^2(k)}{(k^2+g_2(k))^2}\Biggr]\nonumber\\
&&k\frac{\partial}{\partial k}g_6(k)=
-\frac{\hbar k^D N_D}{2}\Biggl[\frac{g_8(k)}{k^2+g_2(k)}
-15\frac{g_4(k)g_6(k)}{(k^2+g_2(k))^2}
+30\frac{g_4^3(k)}{(k^2+g_2(k))^3}\Biggr]\nonumber\\
&&\cdots
\eea

A diagrammatic interpretation of the above equations is straightforward,
provided  one identifies $g_{2n}$ with the $2n$ external legs vertices 
and the denominators $(k^2+g_2(k))^m$ with $m$ propagators joining the 
various vertices; in fact each r.h.s.in Eqs. (\ref{set}) represents
the sum of all the one loop diagrams with fixed number of external legs 
that can be arranged combining the $g_{2n}$. It is then clear that the
above system is an approximation to 
the infinite set of Schwinger-Dyson equations for the Green's functions
at zero external momenta. 
Of course as it is true for the complete Schwinger-Dyson
system, Eq. (\ref{set}) is indeterminate. To make contact with perturbation 
theory, more precisely with the $\hbar$ expansion, we can seek for solutions
where each coupling constant is developed in an $\hbar$-power series
$g_{2n}=g_{2n}^{(0)}+\hbar g_{2n}^{(1)}+\hbar^2 g_{2n}^{(2)}+\cdots$. 
As it is well known 
in this case the solution of the Schwinger-Dyson equations and a fortiori
of our system becomes unique and it may be easily verified that the
ultraviolet behavior of the coupling constants flow at $O(\hbar)$
coincides with the usual one-loop result.
Let us consider for instance the one loop contribution to the perturbative 
$\beta$-function for the $\phi^4$ theory. It is recovered in the 
ultraviolet regime, $k^2 >>g_2$,
by retaining in the r.h.s. the $O(\hbar^0)$ values of the couplings, 
that is their 
boundary values at $k=\Lambda$, $g_4(\Lambda)=\overline{g}_4,~g_6(\Lambda)
=g_8(\Lambda)=\cdots=0$. 
However Eqs. (\ref{set}) show that 
$g_6$ is $O(\overline{g}_4^3)$ 
and then $g_6$ in the r.h.s. of the $\beta$-function of $g_4$ provides a 
$O(\overline{g}_4^3)$ effect. It is then clear that higher loops perturbative
contributions to the $\beta$-function are present in Eqs. (\ref{set}) due 
to the fact that all vertices in the r.h.s. are $k$-dependent. It is
also easy to realize that, in order to recover the full two-loop  
$g_4$ $\beta$-function, it is necessary  to go beyond the LPA
since the $g_6$ and $g_2$ contributions are not sufficient to get 
the complete $O(\overline{g}_4^3)$ effect. 

Going back to Eqs. (\ref{set}), in order to better exploit the 
non-perturbative character of this system, a different kind of truncation is 
needed. A first step toward  its non-perturbative analysis has been taken in 
\cite{ale}. However, due to the appearance of 
non-trivial saddle points in the renormalization group equation studied 
in that paper, the results should be taken quite cautiously. 

The first step beyond the LPA is discussed in the next section.

\section{DERIVATIVE EXPANSION AND ANOMALOUS DIMENSION}

By turning on the scale and  field dependence in $Z_k$ we 
allow for a non-trivial lowest order derivative term in the action 
(\ref{ansatz}), thus obtaining the first improvement to the LPA where it was 
set $Z=1$. In this case we have to deal with two coupled equations for 
$U_k$ and $Z_k$ which can again be reduced to an infinite set 
of coupled equations if one assumes a polynomial expansion for both 
$U_k$ and $Z_k$.

In \cite{alfdar} a specific procedure has been carried out in order to 
determine a differential equation for $Z_k$ in the sharp cut-off limit 
and it has been shown that in $D=4$ for the $O(N)$ symmetric $\phi^4$
theory, the perturbative anomalous dimension of the field to 
$O(\overline{g}_4^2)$ can be obtained from that equation. 
However it was also noticed that, following a different and, in principle,
correct procedure, a different equation for $Z_k$ is obtained. In practice 
the gaussian integration of the fluctuation $\zeta(x)$ in Eq. (\ref{sk})
yields a contribution to $S_{k-\delta k}$ proportional to 
$ln[(\delta^2 S_k/\delta \Phi\delta \Phi)|_{\Phi=\phi}]$ and both 
procedures require to expand the logarithm in order to derive the
differential equation for $Z_k$. In one case, following the steps outlined in 
\cite{fraser} for the one component theory, we split the logarithm 
(again $'$ indicates derivation with respect to the field)
\be \label{fraser}
ln[(\delta^2 S_k/\delta \Phi\delta \Phi)|_{\Phi=\phi}]
=ln \Bigl ( (Z_k)_0 \partial_\mu \partial_\mu
+ (U_k)^{''}_0 + \Delta \Bigr )
\ee
where $(Z_k)_0$ and $(U_k)^{''}_0$ are evaluated at the  constant field 
configuration $\phi_0$ and $\Delta$ is the $x$ dependent part of the 
propagator; the expansion is performed taking $\Delta$ small with respect
to the non-fluctuating part. In the other case, according to 
the procedure developed in \cite{zuk}, 
\be \label{zuk}
ln[(\delta^2 S_k/\delta \Phi\delta \Phi)|_{\Phi=\phi}]=
ln \Bigl ( Z_k k^2 + U_k^{''} + Z_k 
(\partial_\mu \partial_\mu -k^2) +C  \Bigr )
\ee
where $Z_k k^2$ is added and subtracted and the remaining part of 
the propagator $C$ is proportional to derivatives of the field. In the latter 
case the expansion is made requiring that $C$ and the difference
$ Z_k (\partial_\mu \partial_\mu -k^2)$ are small. This is justified due to the
field derivative term in $C$, which is small within the derivative expansion
framework, and to the fact that 
$\partial_\mu \partial_\mu \delta(x-y)$ yields in the infinitesimal 
integration shell $[k-\delta k, k]$ a factor $k^2$ which makes the difference 
in Eq. (\ref{zuk}) vanishing.
It is not surprising that two
expansions obtained for different choices of the expansion parameter,
can be not straightforwardly comparable. As a consequence of the different 
nature of the expansions in Eqs. (\ref{fraser}) and (\ref{zuk}),
we end up with two different equations for $Z_k$. We will come back
to this point in the next Section.

The point of view chosen in \cite{alfdar}
was to test the reliability of these expansions  by comparing the 
values of the anomalous dimension of the field $\eta$ at the lowest 
non-vanishing perturbative order obtained from the equations with the result 
of the usual perturbative method. It turned out that the equation obtained 
from the expansion (\ref{zuk}) provides a value of $\eta$ in agreement 
with perturbation theory in $ D=4$.

In order to support the result of \cite{alfdar} and to show that it is not 
an accidental coincidence, one can easily repeat the calculation of the 
anomalous dimension performed in \cite{alfdar} for the case of the cubic 
theory $\phi^3$ in $D=6$. Obviously eqs. (\ref{ukappa},\ref{set})
are not consistent with the symmetry of the latter theory, but 
the complete equation for the potential (\ref{poteq}) and  the 
equation for $Z_k$, deduced in \cite{alfdar},  are still valid because 
they do not require any assumption on the internal symmetry of $U_k$ and $Z_k$.
The explicit form of the $Z_k$ equation as deduced from the expansion of Eq.
(\ref{zuk}) in $D$ dimensions is ($A= Z_k k^2+U_k^{''}$ and $\hbar=1$)
\be\label{zdimens}
k\frac{\partial}{\partial k}Z_k=-\frac{k^D N_D}{2}\Bigl (
\frac{Z_k^{''}}{A}-
\frac{2 Z_k^{'} A'}{A^2}-
\frac{{Z_k^{'}}^2 k^2}{ D A^2}+
\frac{2 {Z}_k {A'}^2}{3 A^3}+
\frac{8 Z_k^{'} Z_k A' k^2}{3 D A^3}-
\frac{2 Z_k^2 {A'}^2 k^2}{D A^4} \Bigr )
\ee
and, in order to get a perturbative estimate of the anomalous dimension, we 
replace, in the r.h.s. of Eq. (\ref{zdimens}),
$U_k$ and $Z_k$ with the corresponding bare quantities, namely
$\overline{g}_3\phi^3/3!$ and $1$ 
(we perform our calculation in the ultraviolet 
regime where the mass term  in the potential can be neglected). 
The integration of  Eq. (\ref{zdimens}), from $\Lambda$ down to a generic 
value $k<\Lambda$, is then straightforward yielding, in $D=6$,
$Z_k=\overline{g}_3^2/(384 \pi^3) ln(\Lambda/k)$. The anomalous dimension 
is immediately deduced from this $\phi$-independent expression of $Z_k$
\be\label{eta6d}
\eta=-k\frac {\partial}{\partial k} log(Z_k)=\frac{\overline{g}_3^2}{384~\pi^3}
\ee 
and it is in agreement with the perturbative diagrammatic computation
\cite{zinnj}. Actually the usual perturbative approach in this case is very
simple since it involves a one-loop computation and, correspondingly,
we need only one integration of Eq. (\ref{zdimens}) to get the answer, 
differently from the $D=4$ case where, on one side a two-loop diagram is to be
evaluated and, on the other side, a two step integration of Eq. (\ref{zdimens})
is required \cite{alfdar} to determine the lowest (non-vanishing) 
perturbative contribution to  $\eta$. 
However, the above calculation is similar to the one performed for the 
$\phi^4$ theory in four dimensions since in both cases the anomalous dimension
is perturbatively expressed as expansion in powers of the marginal 
dimensionless coupling appearing in the potential.
The running of these couplings with the scale $k$ is only logarithmic 
thus justifying their replacement with the bare constant $\overline{g}_3$
for $D=6$ and $\overline{g}_4$ for $D=4$. The same argument holds
for the field independent part of $Z_k$ which is dimensionless and can be 
replaced with its bare value 1. 

The further important step is to check whether this equation is able to 
reproduce the value of $\eta$ at the non-gaussian (WF) fixed point 
which appears in the scalar theory below four dimensions.

In order to get informations at a fixed point it is convenient to 
express all dimensionful quantities in terms of the running 
scale $k$ and rewrite the coupled equations (\ref{poteq},\ref{zdimens})
in terms of dimensionless variables which can be conveniently introduced 
through the relations $t=ln(k/\Lambda)$, $x=k^{(2-D-\eta)/2}(2/N_D)^{1/2}\phi$,
$u(x,t)=2 k^{-D}U_k/N_D$, $z(x,t)=k^\eta Z_k$.
In these relations powers of $N_D/2$ have also been included in order to get 
a simpler form of the differential equations; indeed the constant $\alpha=
N_D/2$ disappears from Eqs. (\ref{poteq}),(\ref{zdimens}) after the 
replacement $U_k \to \alpha U_k$, $\phi \to \sqrt{\alpha}\phi$,
$Z_k\to Z_k$.
Furthermore, since we are interested in carrying out a numerical analysis of 
the problem, instead of directly attacking Eq. (\ref{poteq}) which contains 
a logarithm, we shall consider the corresponding equation for the 
derivative of the potential, as already performed in \cite{hasen}.
Actually the presence of the logarithm causes a stiffness of our  set of 
equations which is a source  of many numerical drawbacks. 
Therefore, by rearranging the equations in terms of the
derivative of the scaled potential $f(x,t)=\partial u(x,t)/\partial x$ 
we finally get the two coupled differential equations ($a=z+f'$ and this time 
$'$ means derivation with respect to $x$)
\be\label{twoeqs1}
\frac{\partial f}{\partial t}=-D f'+\frac{(D-2+\eta)}{2}(f+xf')-
\frac{z'+f''}{a}
\ee

\be\label{twoeqs2} 
\frac{\partial z}{\partial t}-\eta z-\frac{(D-2+\eta)}{2}x z=-
\Bigl (
\frac{{z''}}{a}-
\frac{2{z'} a'}{a^2}-
\frac{{z'}^2}{ D a^2}+
\frac{2 {z} {a'}^2}{3 a^3}+
\frac{8 {z'} z a'}{3 D a^3}-
\frac{2 z^2 {a'}^2}{D a^4} \Bigr )
\ee
and the fixed points correspond to $t$-independent ($\partial f /\partial t=
\partial z/\partial t=0$) solutions of Eqs. (\ref{twoeqs1},\ref{twoeqs2}).
In the following part of this Section we shall consider only $t$-independent
solutions of Eqs. (\ref{twoeqs1},\ref{twoeqs2}). 

As discussed in \cite{hasen,morris2,morris3}, Eq. (\ref{twoeqs1}) alone allows 
to determine the fixed point structure of the theory and specifically the 
appearance of the WF fixed point is shown for $D<4$. It is 
easy to check that the gaussian fixed point, corresponding to
 $f=0,~z=const\neq 0,~\eta=0$, is a solution of (\ref{twoeqs1},\ref{twoeqs2})
for generic $D$, whereas the determination of the non-gaussian fixed point 
with the corresponding value of $\eta$ requires a numerical analysis.

Eqs. (\ref{twoeqs1},\ref{twoeqs2}) are solved by requiring
the usual normalization of the kinetic term in the action 
and two constraints on $f'$ and $z'$ at $x=0$ which preserve 
the $Z(2)$ symmetry $\phi\to-\phi$ 
\be\label{symmetry}
z(0)=1~~~~~~~~~~~~f'(0)=0~~~~~~~~~~~~z'(0)=0
\ee

As explained in \cite{morris2}, from eqs. (\ref{twoeqs1},\ref{twoeqs2}),
one can easily determine the asymptotic behavior of the solutions
$f(x)$ and $z(x)$ for large $x$, up to two (one for each solution) 
multiplicative constants, $c_f, c_z$. Thus we have enough boundary conditions 
to integrate our equations and to deterimine  the three unknown 
constants $c_f, c_z$ and $\eta$. In order to deal with a
two point boundary problem, we have used, as also suggested in \cite{morris2},
the 'shooting method' embedded in a Newton-Raphson algorithm \cite{numrec}.
The results shown below have been obtained by requiring that the difference
between the constraints in (\ref{symmetry}) and the shooting variables is
less than $\delta=10^{-7}$.

In $D=3$ the value obtained for the anomalous dimension is $\eta=-0.071$,
quite different from the world best determination $\eta=0.035$, 
quoted in \cite{morris2},  or the value determined through the 
$\epsilon$-expansion to the order $O(\epsilon^3)$, $\eta=0.037$
(see \cite{zinnj}). The result is very poor even if compared to the
one obtained starting from the equation for $Z_k$ derived resorting
to a smooth cut-off \cite{morris2}.

However it should be noticed that the situation is less bad if 
$D$ increases. This is illustrated by comparing the value
obtained for $\eta$ at the WF fixed point in $D=3.4$, $\eta=9.74~10^{-3}$
with the estimate of $\eta$ obtained from the
$\epsilon$-expansion \cite{zinnj} to $O(\epsilon^3)$, $\eta=10.70~10^{-3}$,
and $O(\epsilon^4)$,  $\eta=9.62~10^{-3}$. 
In $D=3.6$ we get $\eta=4.30~10^{-3}$ 
to be compared with the value from the $\epsilon$-expansion to 
$O(\epsilon^3)$, $\eta=4.16~10^{-3}$ and to $O(\epsilon^4)$,
$\eta=3.95~10^{-3}$. In $D=3.4$ and $D=3.6$ the differences between the 
various estimates are below $10\%$. In Figs. 1,2,3 we show  
$f(x)$ and $z(x)$ obtained at the WF fixed point for $D=3, 3.4, 3.6$.
 
All curves for $f(x)$ in Fig. 1 intersect the $x$-axis at 
non-vanishing values of $x$,
which correspond to non-zero minima in  the potentials; 
the curves $z(x)$ in  $D=3.4$ and $D=3.6$ are plotted in Fig. 2;
analogously to what is found in \cite{morris2} for the corresponding 
variable in the smooth cut-off framework in $D=3$, $z(x)$ has a maximum
and then decreases for large $x$ (even the dashed curve, although less 
evidently, smoothly decreases after a maximum at about $x=8$).
Conversely, as shown in Fig. 3,
$z(x)$ in $D=3$ is increasing: no maximum has been found even enlarging
the $x$ range to the limits allowed by the integration routine
(in practice the upper limit can be pushed  up to about $x=12$).
Moreover, close to $x=0$,  $z(x)<1$ and the curve has a minimum 
which is not present in the curves in  Fig. 2.

In order to justify this peculiar behavior, we notice that the leading terms
within the brackets in the r.h.s. of Eq. (\ref{twoeqs2}) are the ones 
proportional to $z$ and not $z'$, namely $(2 z {a'}^2)/(3 a^3)-(2 z^2
{a'}^2)/(D a^4)$, that is, as long as $D$ is not close to $3$, these two terms
are dominant since $z>>z'$ but in $D=3$ and in the ultraviolet limit, when 
$a\sim z\sim 1$, they practically cancel out
and the behavior of Eq. (\ref{twoeqs2}) is sensibly modified.

In conclusion, only close to three dimensions, where the infrared effects 
become more and more important, Eqs. (\ref{twoeqs1},\ref{twoeqs2}) 
fail to reproduce the known results about the anomalous dimension.

\section{WH EQUATION AND NON-ANALYTICAL TERMS}

We now move to another point which, in some sense, is preliminary to 
the analysis of the coupled equations for the various parameters entering the 
local action, namely the rise of some undesirable terms as soon as
one goes beyond the LPA in the derivative expansion. The presence
of non-analytical terms has been recognized since long time \cite{sak}in 
relation with the use of a sharp cut-off and recently reconsidered
in \cite{morr,wett3,morris6}. Here, in order to get a clearer
insight into this problem, we shall study the origin of these terms 
working out the various steps 
for the determination of the differential equation for $Z_k$. 

Let us go back to Eq. (\ref{sk}) and (\ref{ansatz}).  
The LPA approximation corresponds to setting
$Z_k=1$, dropping $Y_k$ and all the others higher order derivative 
coefficients and finally restricting to a constant background
$\phi=\phi_0$. In order to get the equation for $Z_k$, we have to release the 
condition $Z_k=1$ and collect all terms proportional to 
$\partial_\mu\phi\partial_\mu\phi$ in the r.h.s. of Eq. (\ref{sk}) 
and, to this purpose, it is necessary to retain a non-constant background 
$\phi(x)=\phi_0+\varphi(x)$. Let us choose for the non-uniform component of 
the background $\varphi(x)$ a single mode with $|q| << k$ 
(the reason for this choice will become apparent later)

\be\label{back}
\varphi(x)= \frac{1}{\sqrt V}\Bigl\{\varphi_q e^{i q\cdot x }+
\varphi_{-q} e^{-i q \cdot x}\Bigr\}
\ee

\noindent
Here $V$ is the volume factor. As before, the fluctuation $\zeta(x)$,
which must be integrated out, contains Fourier modes only in the shell 
$[k-\delta k, k]$  

\be\label{zeta}
\zeta(x)=\frac{1}{\sqrt V}{\sum_{[p]}} \zeta_{p} e^{ip\cdot x}
\ee
\noindent The square bracket in Eq. (\ref{zeta}), $[p]$, is a reminder
that the sum is restricted only to those values of $p$ such that $k-\delta k
\leq |p| \leq k$.

Therefore the original field is split into three parts, 
$\Phi(x)=\phi_0+\varphi(x)+\zeta(x)$. 
In order to pick up contributions up to the second derivative 
term $\partial_\mu\Phi\partial_\mu\Phi$, the functions
$U_k(\phi_0+\varphi(x)+\zeta(x))$ and $Z_k(\phi_0+\varphi(x)+\zeta(x))$ 
must be expanded around $\phi_0$ up to $\varphi^2(x)$, and since the gaussian 
functional integration is exact, as mentioned before, we only need to 
consider terms in $\zeta(x)$ up to $O(\zeta^2(x))$. 
Due to the condition $ q << k$ the linear terms in 
$\zeta(x)$ drop out after the spatial integration. 
Let us consider now the terms proportional to 
$\zeta^2(x)$. To illustrate the various steps made to derive the equation 
and the related problems we do not need to consider here all of them.
It will be sufficient to consider for instance those coming from the 
expansion of the potential. Performing the spatial 
integration with the help of Eqs. (\ref{back},\ref{zeta})
(in the following we shall consider, with no loss of generality
the four dimensional case $D=4$ and, in order to avoid any misunderstanding 
of notations, the $n$ times derived potential is indicated here as $U_k^{(n)}$)
we have

\be\label{int1}
\frac{U^{(2)}_k(\phi_0)}{2} \int d^4x \zeta^2(x) =
\frac{U^{(2)}_k(\phi_0)}{2}{\sum_{[p]}}\zeta_{p}\zeta_{-p}
\ee

\be\label{int2}
\frac{U^{(3)}_k(\phi_0)}{2} \int d^4x \varphi(x) \zeta^2(x) =
\frac{U^{(3)}_k(\phi_0)}{2 \sqrt{V}}\Biggl(
\varphi_q{\sum_{[p]~[-p-q]}} ~~\zeta_{p}\zeta_{-p-q}+
\varphi_{-q}{\sum_{[p]~[-p+q]}} ~~\zeta_{p}\zeta_{-p+q}\Biggr)
\ee

\bea\label{int3}
&&\frac{U^{(4)}_k(\phi_0)}{4} \int d^4x \varphi^2(x) \zeta^2(x) \nonumber \\
&&=\frac{U^{(4)}_k(\phi_0)}{4V}\left (
\varphi^2_q {\sum_{[p]~[-p-2q]}} ~~\zeta_{p}\zeta_{-p-2q}+
2\varphi_q\varphi_{-q} {\sum_{[p]}}~~\zeta_{p}\zeta_{-p}+
\varphi^2_{-q}{\sum_{[p]~[-p+2q]}}~~\zeta_{p}\zeta_{-p+2q}\right )
\eea

One more comment about the notation. 
All sums above are single summations over $p$ and the range spanned by $p$
in each sum is obtained by requiring that all quantities in square brackets
are constrained into the shell $[k-\delta k, k]$. For instance in the 
first  sum in Eq. (\ref{int2}), $[p]~[-p-q]$ indicates the double constraint
$(k-\delta k)\leq |p|, |-p-q| \leq k$, which in turn determines the values
of $p$ selected in the summation. Then, as soon as we turn on a non-uniform 
background $\varphi(x)$, we end up with a deformation of the integration 
region. The new integration region for $p$ is now given by the overlap of 
two shells with one of the two  depending on $q$. 
Only at $q=0$ the two shells are both reduced  to the original one. 
The reason for the appearance of the non-analitycal terms has to be traced 
back to this boundary effect.

In order to deal with shorter expressions we limit ourselves to the case of a 
$\phi$-independent $Z_k$. Then, with the help of 
Eqs. (\ref{int1},\ref{int2},\ref{int3}), we can easily expand the action 
$S_k[\phi_0+\varphi+\zeta]$ around $\phi_0$ and perform the 
quadratic integration in $\zeta$ obtaining ($\hbar=1$)

\bea
&&exp \left\{ {-S_{k-\delta k}[\phi_0 + \varphi]} \right\}
= exp\left \{ {-S_k[\phi_0 + \varphi]} \right \} 
~exp \left\{ {\frac{1}{2}{\sum_{[p]}}~~ ln G^{-1}(p)}\right \}\nonumber\\ 
&&\times \Biggl\{ 1 + \frac{{U^{(3)}_k(\phi_0)}^2}{4 V} 
\varphi_q\varphi_{-q}\Biggl [{\sum_{[p]~[p+q]}} 
G(p)G(p+q) + {\sum_{[p]~[p-q]}} G(p)G(p-q)\Biggr ] \Biggr \} 
\eea
 
\noindent
where we have introduced the propagator-like  notation  
$G^{-1}(p) = Z_k p^2 + U^{(2)}_k(\phi_0) + U^{(4)}_k(\phi_0)
\varphi_q\varphi_{-q}/V$.

Since we are just interested in the $Z_k$
evolution, we again neglect cubic and higher powers of $\varphi$
and expand $G(p)$, $G(p+q)$, $G(p-q)$ around $q=0$ up to $O(q^2)$,
obtaining ($A(p)=Z_k k^2+U_k^{(2)}(\phi_0)$ )

\bea\label{azione}
S_{k-\delta k}[\phi_0 + \varphi] =&& S_k[\phi_0 + \varphi] 
+ \frac{1}{2}{\sum_{[p]}}~ln \Bigl ( A(p)\Bigr)
+\frac{U^{(4)}_k(\phi_0)}{2 V}\varphi_q\varphi_{-q}
{\sum_{[p]}}\frac{1}{A(p)}\nonumber\\
&&-\frac{{{U_k}^{(3)}(\phi_0)}^2}{4 V}\varphi_q\varphi_{-q}\Biggl\{
 {\sum_{[p]~[p+q]}}\frac{1}{A^2(p)}\Bigl[
1-\frac{q^2 Z_k}{A(p)}+\frac{4 (p\cdot q)^2 Z_k^2}{A^2(p)}\Bigr]
\nonumber\\
&&+{\sum_{[p]~[p-q]}} \frac{1}{A^2(p)}\Bigl[
1-\frac{q^2 Z_k}{A(p)}+\frac{4 (p\cdot q)^2 Z_k^2}{A^2(p)}\Bigr]
\Biggr\}
\eea

At this point if  we just neglect the boundary problem and everywhere 
integrate the momentum $p$ within the  shell $[k-\delta k,k]$, from the 
terms in Eq. (\ref{azione}) not depending on $\varphi$ we get 

\be\label{poteqz}
k\frac{\partial}{\partial k}U_k(\phi_0)=-
\frac{k^4}{ 16 \pi^2}ln(Z_k k^2+U_k^{''}(\phi_0))
\ee
This  equation replaces  Eq. (\ref{poteq}) in $D=4$.
From the terms proportional to  $\varphi_q\varphi_{-q}$, we get 
the second derivative with respect to $\phi_0$ of Eq. (\ref{poteqz}).
This is simply because those terms come from an expansion 
of Eq. (\ref{poteqz}) evaluated at $\phi(x)=\phi_0 + \varphi(x)$
around $\phi_0$.

Finally if we collect the terms proportional to $q^2 \varphi_q\varphi_{-q}$
we get the evolution equation for $Z_k$

\be\label{zequ}
k\frac{\partial}{\partial k}Z_k=-
\frac{ k^4}{ 16\pi^2}\Biggl (
\frac{Z_k {U_k^{(3)}(\phi_0)}^2}{A(k)^3}
-\frac{{Z_k^2 U_k^{(3)}(\phi_0)}^2 k^2}{A(k)^4}
\Biggr )
\ee

In the absence of the boundary problem and under the assumption of a field 
independent $Z_k$, (\ref{poteqz}) and (\ref{zequ}) are the system of coupled 
differential equations for $U_k$ and $Z_k$.
In the more general case of a field dependent $Z_k$ by following the same 
steps as before we find that the equation that replaces (\ref{zequ}) is

\be\label{zfras}
k\frac{\partial}{\partial k}Z_k=-\frac{k^4 }{16 \pi^2}\Bigl (
\frac{Z_k^{''}}{A}-
\frac{2 Z_k^{'} A'}{A^2}-
\frac{{Z_k^{'}}^2 k^2}{ 4 A^2}+
\frac{{Z}_k {A'}^2}{ A^3}+
\frac{Z_k^{'} Z_k A' k^2}{ A^3}-
\frac{Z_k^2 {A'}^2 k^2}{A^4} \Bigr )
\ee

Several comments are in order.
This equation has already been derived by following different steps in 
\cite{alfdar2,alxtesi}. As already noticed in \cite{alfdar} it 
is substantially different from Eq.(\ref{zdimens}), namely both equations
contain the same kind of terms but with different numerical coefficients. 
In the previous Section we have already explained that to derive Eq.
(\ref{zdimens}), an expansion where the "kinetic" term
$ Z_k (\partial_\mu \partial_\mu -k^2$) is considered small w.r. to 
the "potential" term $Z_k k^2 + U_k^{''}$ is used. The expansion that leads to 
Eq.(\ref{zfras}) is of a completely different nature. In this latter case
the potential term is considered small w.r. to the kinetic term. 
It is then not surprising that we get different results. It is well 
known that no one of these expansions can be considered as 
definitely superior to the other. The choice of one of them is 
rather dictated by the physical problem under investigation.

It may be stressed at this point that the fact of having derived two 
evolution equations for $Z_k$, namely Eqs. (\ref{zfras}) and (\ref{zdimens}), 
is not the consequence of having used two 
different definitions of this parameter (which is actually introduced  
by the derivative expansion (\ref{ansatz}) of the action whose evolution 
is determined by the integration in Eq. (\ref{sk}), with the boundary 
condition $Z_\Lambda=1$). Rather, it is the consequence of having used 
two different approximations, which in turn are necessary since the full 
problem cannot be solved exactly. More specifically, as we have just seen,
we need to insert explicitly a non-uniform
( and slowly varying, so that higher derivatives terms in the derivative 
expansion can be safely neglected) background field to read 
the differential evolution equation for the coefficient $Z_k$.
Now the fluctuation operator ~$\frac{\delta^2 S_k}{\delta\phi(x)\delta
\phi(y)}$~ is obviously no longer diagonal in momentum space. 
To determine then a differential equation for $Z_k$,
{\it we need a further expansion} in which a piece of the propagator is 
considered small with respect to the rest of the propagator. 
In the derivation of Eqs.(\ref{zdimens}) and (\ref{zfras}) above 
this additional expansions correspond respectively to the physical 
case where the kinetic energy is small with respect to the potential 
energy and viceversa. 

The method exposed above to obtain Eq. (\ref{zdimens}) is different
from the methods used respectively in  \cite{alfdar2} and \cite{alxtesi}.
Nevertheless
the expansion adopted in the three cases, namely a small coupling expansion, 
is the same and this explains why these methods all lead to the same equation.

Finally Eq. (\ref{zfras}) is obtained by neglecting the 
boundary effects due to the deformation of the integration shell.
Let us now go back to Eq. (\ref{azione}) and work out explicitly the 
consequences of this boundary distortion.
To illustrate this problem we just need to consider one term. 
Let us take for example the term 

\be\label{utre}
\frac{{{U_k}^{(3)}(\phi_0)}^2}{4 V}\varphi_q\varphi_{-q}
{\sum_{[p]~[p+q]}}\frac{1}{A^2(p)}
\frac{Z_k q^2}{A(p)}
\ee

To further simplify the point, let us write down explicitly the above 
indicated  sum in $D=1$ dimensions. The only additional complications 
that would appear when considering higher dimensions are some factors
coming from the angular integrations. By trivially replacing 
the sums with the integrals we get as a result of the overlap of the 
two shells ($L$ is the volume of the system)

\bea\label{nonan}
{\sum_{[p]~[p+q]}}\frac{1}{A^2(p)}
\frac{Z_k q^2}{A(p)}&&= 
Z_k q^2 \frac{L}{2\pi}\Biggl\{
\int^{-k +\delta k}_{-k+q} \frac{dp}{{(Z_k p^2+U_k^{(2)}(\Phi))}^3}+  
\int^{k}_{k - \delta k +q} \frac{dp}{{(Z_k p^2+U_k^{(2)}(\Phi))}^3}
\Biggr\}\nonumber\\
&&=Z_k q^2 \frac{L}{2\pi}\frac{2}{{(Z_k k^2+U_k^{(2)}(\Phi))}^3}
\Bigl(\delta k - q\Bigr)
\eea

\noindent 
It must be noted that the companion term in eq.(\ref{azione}), which differs
from the one in Eq. (\ref{nonan}) only for the integration limits, 
gives the same result and we do not 
get any cancellation of the term proportional to $q$.
The above result explains our previous statement that the appearance of 
non analytic terms in $q^2$ is simply due to the distortion of the
integration domain induced by the non uniform background configuration.

In Eq. (\ref{nonan}) a term proportional to $q$ appears {\it which is not 
proportional to $\delta k$}. This in turn seems to imply that the very 
possibility to establish a differential equation for $Z_k$
is in danger and actually  the presence of such terms has been considered 
as a serious drawback that blocks the way to any application of the WH method 
beyond the LPA approximation. That consideration has convinced some authors
that to overcome this problem a smooth cut-off rather than a sharp cut-off 
should be used. We shall come back to the smooth cut-off later. 

The above derivation actually teaches us that there is no serious drawback 
as far as we correctly interpret our equations. The undesired non-analytic 
terms that appear on the r.h.s. of the defining equation for 
$S_{k-\delta k}$ and that are not contained in the original derivative
expansion ansatz, always appear in combination with 
$\delta k$ as in the example above. This means that as long as we keep 
$q << \delta k$, that is as long as the non-uniformity of the background field
is small compared to the width of the shell within which 
the modes are treated as independent, these terms can be coherently neglected 
and the equations above are the correct WH RG equations in this approximation.
This is actually one important result of our analysis. 
The very possibility to implement the RG trasformations 
as a system of differential equations for $U_k$ and $Z_k$ ( i.e. to consider 
such a truncation to the infinite system of equations  for all the 
coefficient functions  that appear in the derivative expansion as a good 
approximation) is intimately related to the nature of the background field 
$\phi(x)$ that we consider. It has to be sufficiently smooth over the scale 
$\delta k$ otherwise the differential equations simply cease to be valid. 
The important point here is that not only we learn that it is possible 
and perfectly legitimate to establish differential equations that  
implement the infinitesimal RG transformation even beyond the LPA 
approximation, but we also obtain the limit of validity of these equations.

It is worth at this point to pause for a moment and review all the steps 
and approximations involved in the above derivation. First note that the 
wilsonian renormalization group transformation gives:

\be\label{aux}
S_{k-\delta k}(\phi)= S_{k-\delta k}(\phi) + \delta k (\cdots) 
+ O((\delta k)^2)
\ee

From the derivation of Eq.(\ref{aux}) it is clear that $\delta k$ is the
momentum range above which the modes are considered as independent, that is
any momentum function $f(p)$ that appears in the above equation is 
practically constant within this range. This in turn means that we can 
neglect the $O((\delta k)^2)$ terms or, in other words,
that the above equation can be rewritten as a differential equation.
This is what is meant when the above equation is referred as an "exact 
differential renormalization group equation" and this is also the meaning 
of the equivalent statement saying that the gaussian integration is "exact". 
Actually all that means that the $O((\delta k)^2)$ terms can be consistently 
neglected. 

\noindent 
Next we approximate $S_k$ by a derivative expansion. We have 
seen that as soon as we go beyond the lowest order and allow for a non
constant background field we encounter singularities that make the formal
mathematical limit $\delta k\to 0$ ill defined. We have seen above that 
this is none of a problem as far as the scale at which singular terms 
appear is kept far from the "resolution scale" ~$\delta k$. 
This is nothing but the condition $ q << \delta k$. 

\noindent 
The two points above can be easily understood by considering 
the following hydrodynamical example. 
To give a differential form to the equation 
for, say, the density $\rho$ of a fluid we need two conditions 
to be satisfied. The infinitesimal volume $d^3x$ has to be "sufficiently 
small" so that  the density as well as any other macroscopic quantity can 
be considered as constant within this volume. At the same 
time it has to be "sufficiently large" so that a macroscopic number of 
molecules is contained within this volume. 
Under the above conditions we can write the evolution equation 
for the density in a differential form because the $O((d^3x)^2)$ 
contributions can be neglected and because due to the second condition 
we can neglect the singularities coming from the molecular scale ~$a$~, once 
our "resolution scale" ~$dx$~  has been taken much bigger than ~$a$~ : 
$dx >> a$. The complete analogy between the approximations involved in the
derivation of our equations and the hydrodynamical example should be, by now,
clear. 
Obviously the differential equation for $\rho$ obtained this way becomes 
less reliable and ultimately totally wrong when we want to 
describe physical phenomena whose resolution scale approaches
the molecular scale $a$. The same warning applies to our case when the 
background field is not sufficiently smooth within $\delta k$, and this 
happens when $q \to \delta k$. 

Let us now turn our attention to the smooth cut-off procedure.
To illustrate the point we can again reconsider the previous  example and
implement the constraint in Eq.(\ref{utre}) by means of 
differences between theta functions in the following way

\be\label{theta}
{\sum_{[p]~[p+q]}}\frac{1}{A^2(p)}
\frac{Z_k q^2}{A(p)}= 
Z_k q^2 \frac{L}{2\pi}
\int^{+\infty}_{-\infty}dp~~
\frac{\Theta_0 (p^2, k^2, (k-\delta k)^2)
\Theta_0 ((p+q)^2, k^2, (k-\delta k)^2)}
{{(Z_k p^2+U_k^{(2)}(\phi_0))}^3}
\ee

\noindent
where we have defined 
$\Theta_0 (p^2, k^2, (k-\delta k)^2)= 
\theta (p^2-(k-\delta k)^2)-\theta(p^2-k^2)$.

The effect of a smooth cut-off is implemented here by replacing the 
$\Theta_0 (p^2, k^2, (k-\delta k)^2)$ function with a smoothened version
$\Theta_ \epsilon(p^2, k^2, (k-\delta k)^2)$ where $\epsilon$ has the 
dimension of a momentum and is a new cut-off that we introduce in the theory;
in the limit $\epsilon\to 0$ it should give  $\Theta_0$. 
It is easy to see that all the non-analytic terms disappear.
In fact by expanding the function 
$\Theta_\epsilon((p+q)^2, k^2, (k-\delta k)^2)$
around $p^2$  we get (from now on we omit $k$ and $k-\delta k$ 
in the argument of the $\Theta_{\epsilon}$ function)

\be
\Theta_\epsilon((p+q)^2)=\Theta_\epsilon(p^2)
+\frac{d \Theta_{\epsilon} (p^2)}{d p^2}
\Bigl (2pq+ q^2 \Bigr )
+ \frac{1}{2}\frac{d^2 \Theta_{\epsilon} (p^2)}{d (p^2)^2}
\Bigl (2pq+ q^2 \Bigr )^2 +\cdots
\ee

Being now the integration region symmetric all contributions from 
odd powers of $p$ vanish and consequentely odd powers of $q$ are absent.
As we shall see in a moment we have gained nothing from having made the
non-analytic terms disappear. Let us focus on the 
function  $\Theta_{\epsilon}$ and regard it as a function of $|p|$.
The scale
$\epsilon$ gives the size of the region over which $\Theta_{\epsilon}$ 
as a function of $|p|$ changes significantly from zero to one. This 
means that its first derivative with respect to $|p|$ is 
$O(\frac{1}{\epsilon})$ in the 
two regions around $k$ and $k-\delta k$, and zero everywhere else.
Analogously its second derivative is $O(\frac{1}{\epsilon^2})$ in the 
same region and zero everywhere else, and so on.

We could expect that  $\epsilon$ is introduced just as an intermediate 
step and that the final results are obtained after sending $\epsilon$ 
to zero and they are finite. The above result shows that this is not the case. 
As a consequence of the introduction of the smoothening cut-off
$\epsilon$ we have generated $\frac{1}{\epsilon}$ divergences which have 
already been noticed in \cite{morr,wett3,morris6}.
This could appear at a first sight as a disaster and again we might wonder 
whether it is possible to establish
differential equations for $U_k$ and $Z_k$ ( and more generally for the 
coefficient functions of the derivative expansion)  under these conditions.
Actually we can repeat here the same kind of considerations that we have 
done in connection with the appearance of the non analytic terms. 
After performing the integration in the momentum $p$ we see that 

i) the terms proportional  $\Theta_\epsilon(p^2)$ give contributions 
O($\delta k$). This is because 
the integration region where the integrand is significantly different
from zero is the shell $[k-\delta k, k ]$;

ii) the terms proportional to  $\frac{d \Theta_{\epsilon}(p^2)}{d p^2}$ 
give contributions proportional to $\Bigl (\frac{q}{k}\Bigr ) q$, 
$\Bigl (\frac{q}{k}\Bigr )^3 q$, and so on. This is because the region over
which the integrand is significantly different from zero has now a width of
$\epsilon$ and we get then a factor $\epsilon$ coming from the width and 
a factor $\frac{1}{\epsilon}$ coming from the derivative;

iii) the terms proportional to  $\frac{d^2 \Theta_{\epsilon}(p^2)}{d (p^2)^2}$ 
give contributions proportional to
$\Bigl(\frac{q}{\epsilon}\Bigr) q $,
$\Bigl(\frac{q}{k}\Bigr)^2\Bigl(\frac{q}{\epsilon}\Bigr) q $, 
$\Bigl(\frac{q}{k}\Bigr)^4\Bigl(\frac{q}{\epsilon}\Bigr) q $,
$\Bigl(\frac{q}{k}\Bigr)^6\Bigl(\frac{q}{\epsilon}\Bigr) q $, and so on. 
This is because the region over which the integrand is significantly 
different from zero has a width $\epsilon$ as before but this time  
we get a contribution $O(\frac{1}{\epsilon^2})$ from the second derivative of 
$\Theta_{\epsilon}$.  

Let us now comment on the above results. First we make the choice
$\epsilon \sim q$, that is a very good choice 
for $\epsilon$ once we remember its role of smoothening parameter 
around the points $k$ and $k - \delta k$ and that we want 
to integrate essentially only the modes within the shell.
It is now apparent that as long as we keep the scale  $q<<\delta k$  
the dominant contributions come from 
the terms proportional to $\Theta_\epsilon(p^2)$  and all the contributions 
coming from the derivatives of $\Theta_\epsilon(p^2)$ can be coherently 
neglected and the final output is practically equivalent to the sharp cut-off
one.

An important comment has to be made at this point. The particular choice
of the smoothening function $\Theta_\epsilon$
that we have done above, more precisely the
fact that we have chosen it to be an even function of $|p|$, is responsible 
for getting only even powers of $q$ in the above results, in other words we 
do not get non analytic terms in $q^2$. But this is absolutely irrelevant 
because what matters to establish the differential equations is the condition
$q<<\delta k$ irrespectively of the fact that we have odd or even powers
of $q$. In fact had we chosen $\Theta_\epsilon$ to have an odd 
dependence on $|p|$ we would have also found  odd powers of $q$. 
Nevertheless under the conditions $q<<\delta k$ and $q<< k$,  
we can equally well neglect all the terms apart from the first and again 
recover the differential equations. 

What we have just seen proves that actually there is no conflict between the 
sharp and the smooth cut-off approach. These two procedures, once correctly
implemented, give precisely the same results as it should be expected.
We have also learned under which conditions the couple of differential 
equations for $U_k$ and $Z_k$ are valid.

Another additional comment peculiar to the smooth cut-off procedure has to 
be made.
Due to the disappearance of the non-analytic terms there has been 
in recent years some preference for the smooth cut-off implementation of the 
differential RG transformations w.r. to the sharp cut-off. 
As we can easily see  from the points (i) (ii) 
and (iii) a large value of $\epsilon$ looks even more efficient in 
suppressing the undesired terms and we could be led to the conclusion 
that we can accomodate such large values of $\epsilon$ within the smooth 
cut-off approach. Some results have been recently derived by making use
of this apparent better flexibility of the smooth cut-off versus the 
sharp cut-off. Actually the condition of having a small $\epsilon$ 
( where small means not too big compared to $\delta k$) 
is necessary to be coherent with the very strategy of the 
RG method as exposed before. Substantially only the modes within the shell 
have to be integrated out. By considering larger values of $\epsilon$ 
we move toward the independent mode approximation. The conclusion is that
those results that have been obtained within the framework of the 
smooth cut-off procedure with the help of cut-off functions whose 
typical width is of $O(k)$, where $k$ is the UV cut-off, 
have to be taken with a grain of salt and the improvement w.r. to
the perturbative result (independent mode approximation) is not very 
much under control.

\section{summary and conclusions}

We have carefully analysed the derivative expansion for the 
effective action and shown how the exact renormalization group
equations for the coefficient functions $U_k$ and $Z_k$ are obtained. 
The most important lesson we have learned is that the system of coupled 
differential equations for them can actually be established
provided the background field around which the quantum fluctuations are
integrated is sufficiently smooth. The width $\delta k$ in Fourier space 
measures the momentum range within which the modes are treated
as independent, meaning that any function $f(p)$ that has to be integrated 
in the shell $[k-\delta k, k]$ is considered to be  constant within it. 
The key point of the method is that this shell has to be on the one hand 
sufficiently small, i.e. $\delta k << k$, where $k$ is the UV 
cut-off, so that the feed-back of the higher energy modes on the lower 
ones is correctly taken into account.
On the other hand, and this is the crucial point that we want to emphasize
here, it has to be sufficiently large in such a way that within this range the
background field can be considered practically flat, i.e. $q << \delta k$. 
Having made clear this point we have seen that the non-analytic terms are no 
more source of problems and the differential equations
can be safely established under these conditions. We have also proven that
the introduction of a smooth cut-off practically does not change our 
conclusions. In fact, under the conditions quoted above, sharp and smooth
cut-off produce precisely the same results. 
A word of caution has to be said regarding the results 
that  have been obtained within the smooth cut-off procedure by allowing
the smoothening scale $\epsilon$ to take values $O(k)$. 
In this case the danger is that the width of the shell $\delta k$ gets 
infinitesimal w.r. to the smoothening scale which means that the modes
are practically all treated as independent and then there is no control
on the possible improvement on the perturbative results.

Another important point we have learned from the above analysis is that 
depending on the physical problem at hand different expansions can be 
envisaged and this lead to different equations. The 
typical situations illustrated are the one in which the kinetic
energy term is considered small compared to the potential and the opposite
one. These two cases  respectively produce  Eq.(\ref{zdimens})
and Eq.(\ref{zfras}) above. 

One third important result of our analysis is that ( see Sec. III) the
study of the coupled system of equations for $U_k$ and $Z_k$ 
has shown that we are unable to reproduce the value of $\eta$ at the 
Wilson-Fisher fixed point in $D=3$ dimensions. 
Our previous results teach us that the reason of this failure is not to be 
searched in an 
intrinsic weakness of the sharp cut-off procedure versus the smooth 
cut-off. Presumably the reason is that we enter a region where the equations
themselves are no longer valid. An indication for that can be seen
in the behaviour of $Z_k$. When it deviates from  $Z_k=1 $, then 
less smooth configurations play an important role. 
We may expect that a better result could be obtained once
some other coefficient functions of the derivative expansion are added 
and the new set of coupled differential equations derived.

\acknowledgments

We would like to thank J. Alexandre, M. Reuter and J. Polonyi for helpful 
discussions.

\vfill\eject

\begin{figure}
\psfig{figure=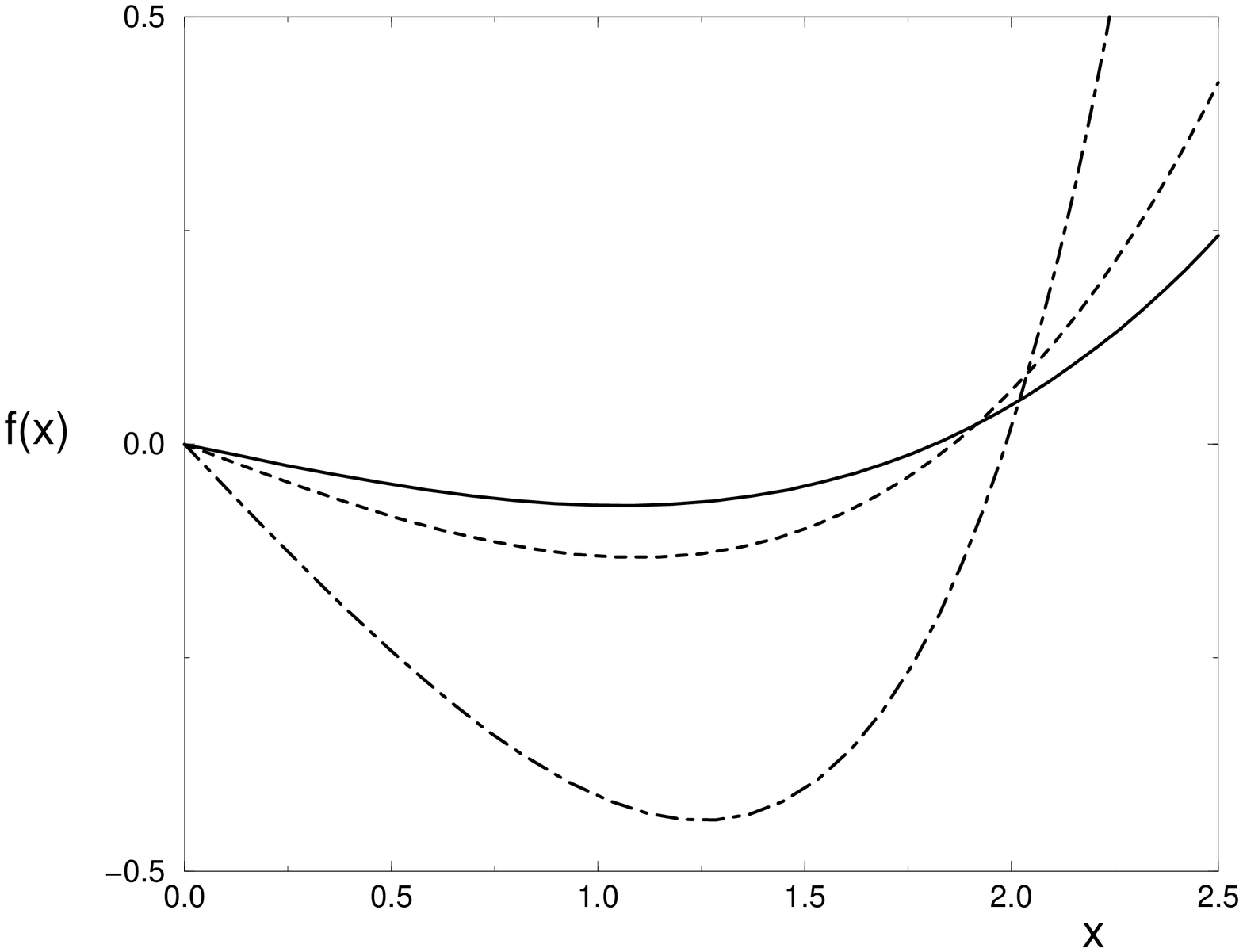,height=12.5 true cm,width=15 true cm,angle=270}
\caption{The derivative of the WF fixed point potentials $f(x)$, in 
$D=3$ (dot-dashed line), $D=3.4$ (dashed line), $D=3.6$ (solid line).}
\end{figure}

\vfill\eject

\begin{figure}
\psfig{figure=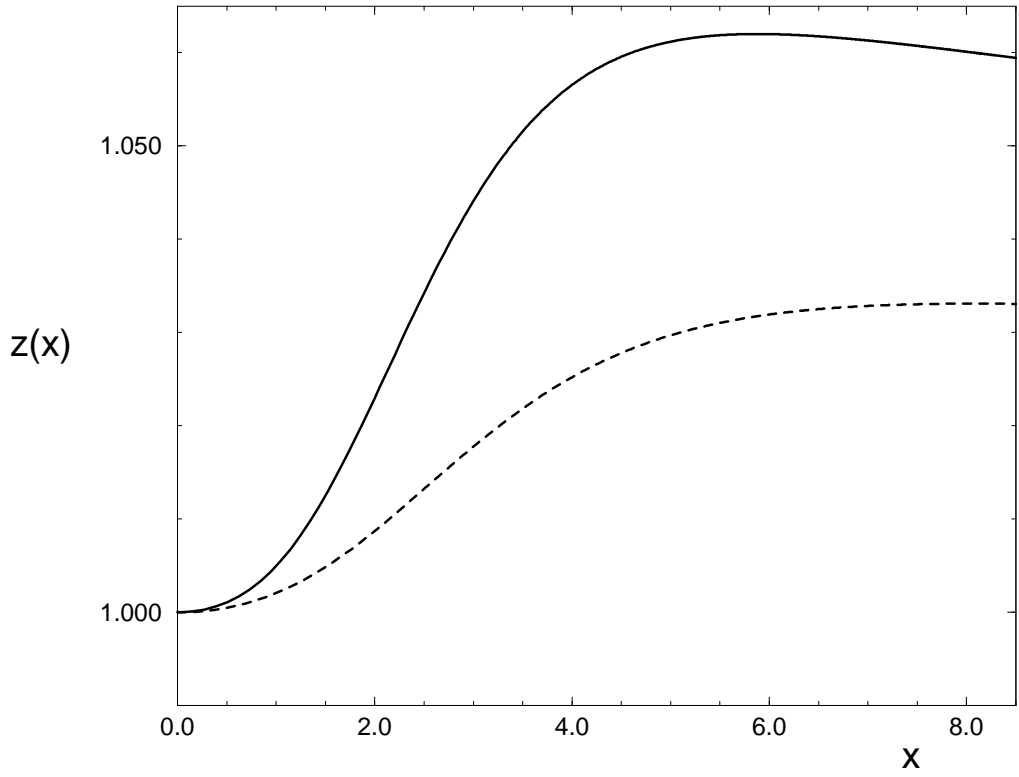,height=12.5 true cm,width=15 true cm,angle=270}
\caption{$z(x)$,  at the WF fixed point in  $D=3.4$ (solid line), 
$D=3.6$ (dashed line).}
\end{figure}

\vfill\eject

\begin{figure}
\psfig{figure=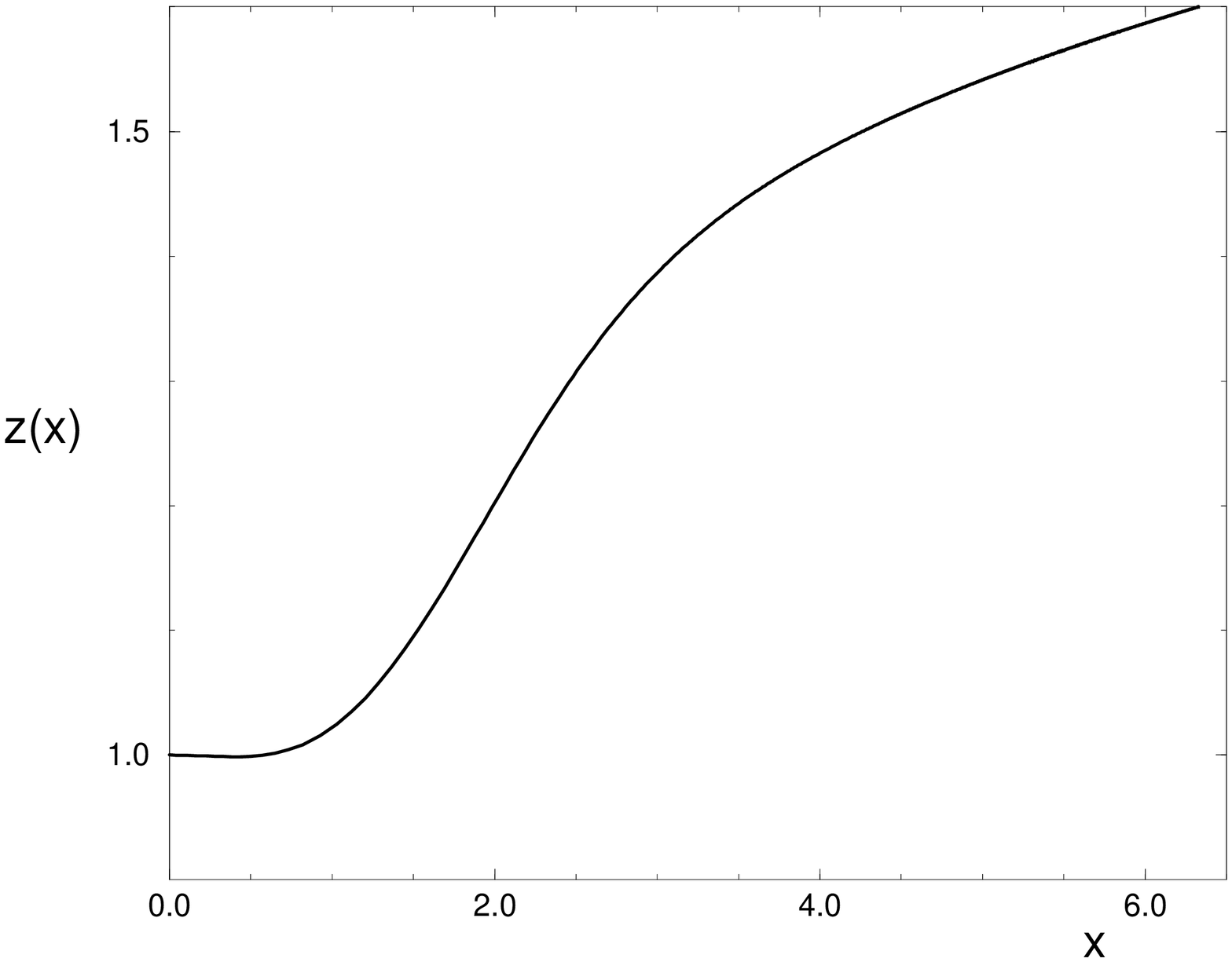,height=12.5 true cm,width=15 true cm,angle=270}
\caption{$z(x)$,  at the WF fixed point in $D=3$.}
\end{figure}


\begin{thebibliography}{12}
\bibitem{wilson} L. P. Kadanoff, Physica {\bf 2}, 263 (1966);
K. Wilson, Phys. Rev. {\bf B4}, 3184 (1971); 
K. Wilson and J. Kogut, Phys. Rep. {\bf 12}, 75 (1974). 
\bibitem{wegner} F. J. Wegner and A. Houghton, Phys. Rev {\bf A8}, 401 (1972).
\bibitem{polch} J. Polchinski, Nucl. Phys. {\bf B 231}, 269 (1984). 
\bibitem{hasen} A. Hasenfratz and P. Hasenfratz, Nucl Phys. {\bf B270}, 
685 (1986).
\bibitem{wett} C. Wetterich, Nucl. Phys. {\bf B352}, 529 (1991).
\bibitem{wett1} C. Wetterich, Phys. Lett. {\bf B 301}, 90  (1993).
\bibitem{morr} T. R. Morris, Int. J. Mod. Phys. {\bf A 9}, 2411 (1994). 
\bibitem{morris2} T. R. Morris, Phys. Lett. {\bf B 329}, 241 (1994).
\bibitem{wett2} A. Ringwald and C. Wetterich, Nucl.Phys. {\bf B 334}, 
506  (1990). 
\bibitem{bran1}J.Alexandre, V.Branchina and J.Polonyi,
``Instability induced renormalization'', e-Print Archive: cond-mat/9803007.
\bibitem{boh} T. Papenbrock and C. Wetterich, Z. Phys. {\bf C 65}, 519 (1995).
\bibitem{alfdar}A. Bonanno and D. Zappal\`a, Phys. Rev. {\bf D 57}, 
7383 (1998). 
\bibitem{jean} See \cite{bran1} for the more  general case when the saddle 
point of the gaussian integral is not the trivial one $\zeta=0$ as it is 
implicitly supposed in the present text.
\bibitem{nicol} J.F. Nicoll, T.S. Chang and H.E. Stanley, Phys. Rev. Lett.  
{\bf 33}, 540 (1974).
\bibitem{ale}J.Alexandre, V.Branchina and J.Polonyi, Phys. Rev. {\bf D 58 }: 
016002 (1998). 
\bibitem{fraser} I. Aitchison and C. M. Fraser, Phys. Lett. 
{\bf B 146}, 63 (1984); Phys. Rev.  {\bf D 31}, 2605 (1985); C. M. Fraser,
Z. Phys. {\bf C 28}, 101 (1985).
\bibitem{zuk} J. A. Zuk, J. Phys. {\bf A 18},1795 (1985); Phys. Rev. 
{\bf D 32}, 2653 (1985).
\bibitem{zinnj} J. Zinn-Justin, ``Quantum Field Theory and Critical 
Phenomena''
Oxford Science Publications, (1990), Clarendon Press Oxford.
\bibitem{morris3} T. R. Morris, in {\it RG96},  e-Print Archive: 
hep-th/9610012.
\bibitem{numrec} W. H. Press {\it et al.}, ``Numerical Recipes, the art of 
scientific computing'', 2nd edition (1992), Cambridge University Press.
\bibitem{sak} J. Sak, Phys. Rev. {\bf B 8},281 (1973).
\bibitem{wett3} C. Wetterich, Z. Phys. {\bf C 57}, 451 (1993).
\bibitem{morris6} T. R. Morris, Nucl. Phys. {\bf B 458 [FS]}, 477 (1996).
\bibitem{alfdar2} A. Bonanno and D. Zappal\`a, Phys. Rev. {\bf D 55 }, 
6135 (1997); Phys. Rev. {\bf D 56}, 3759 (1997).
\bibitem{alxtesi} J. Alexandre, PhD thesis, Louis Pasteur University,
Strasbourg, "Renormalisation en Presence de Condensat".



\end{thebibliography}
\end{document}